\documentclass[12pt]{article}
\usepackage{epsfig}
\def\beq{\begin{equation}}
\def\eeq{\end{equation}}
\def\bea{\begin{eqnarray}}
\def\eea{\end{eqnarray}}
\def\bq{\begin{quote}}
\def\eq{\end{quote}}

\def\NP{{\it Nucl.Phys.} }
\def\PL{{\it Phys.Lett.} }
\def\PR{{\it Phys.Rev.} }

\def\gappeq{\mathrel{\rlap {\raise.5ex\hbox{$>$}}
{\lower.5ex\hbox{$\sim$}}}}

\def\lappeq{\mathrel{\rlap{\raise.5ex\hbox{$<$}}
{\lower.5ex\hbox{$\sim$}}}}

\def\Toprel#1\over#2{\mathrel{\mathop{#2}\limits^{#1}}}

\begin{document}
\pagestyle{empty}
%\begin{flushright}
%{CERN-TH/2001-167}
%hep-ph/th number??\\
%\end{flushright}
\vspace*{5mm}
\begin{center}
{\bf DIFFUSION CORRECTIONS TO THE HARD POMERON\footnote{Work supported in
part by EU QCDNET Contract FMRX-CT98-0194, by MURST (Italy),
and by the US Department of Energy.}}
\\
\vspace*{1cm}
{\bf Marcello CIAFALONI}\footnote{On sabbatical leave from Dipartimento di
Fisica and INFN, Firenze, Italy.}\\ \vspace{0.3cm}
Theoretical Physics Division, CERN\\
CH 1211 Geneva 23\\ \vspace{0.5cm}
{\bf Martina TAIUTI}\\ \vspace{0.3cm}
Dipartimento di Fisica dell'Universit\`a, Firenze, Italy\\ \vspace{0.5cm}
and\\ \vspace{0.5cm}
{\bf A. H. MUELLER}\footnote{On sabbatical leave from Columbia
University,
New York, U.S.A.}\\ \vspace{0.3cm}
LPT, Universit\'e de Paris-Sud, 91405 Orsay, France

\vspace*{1cm}

{\bf Abstract} \\ \end{center}

\vspace*{5mm}

The high energy behaviour of two-scale hard processes is investigated in the
framework of small-$x$ models with running coupling, having the Airy
diffusion model as prototype. We show that, in some intermediate high energy
regime, the  perturbative hard Pomeron exponent determines the energy
dependence, and we prove that diffusion corrections have the form hinted at
before in particular cases. We also discuss the breakdown of such regime at
very large energies, and the onset of the non-perturbative Pomeron
behaviour.

\vspace*{3cm}

%\begin{flushleft} CERN-TH/2001-167 \\
%25 June 2001
%\end{flushleft}
\newpage
%\pagestyle{empty}
%\clearpage\mbox{}\clearpage

\setcounter{page}{1}
\pagestyle{plain}

\section{Introduction and Outline}

High energy hard scattering has received considerable attention in recent
years. The essential problem is to determine the Green's function
$G(Y;k,k_0)$ for gluon--gluon forward scattering, where $k$ and $k_0$ are the 
mass
scales of the gluons and $Y=\log(s/kk_0)$ is the rapidity corresponding to a
center-of-mass energy squared $s$. The classic calculations done by Balitsky,
Fadin, Kuraev and Lipatov (BFKL) \cite{bfkl} several years ago corresponds to
an approximation (the leading series of powers in $\alpha Y$) where the QCD
running coupling $\alpha_s(t)$ is treated as a constant, $\alpha$. In this
case the high-energy behaviour of $G$ is determined by the rightmost
singularity in the $\omega$ plane, where $\omega$ is the variable conjugate to
$Y$. This singularity at $\omega_s = \bar{\alpha}\chi_m = {\alpha
  N_c\over\pi}\chi_m$ is given in terms of the saddle point $\chi_m$ of the
function $\chi(\gamma)$ which gives the eigenvalues of the BFKL kernel.

When higher order corrections \cite{fl,cc1} to the BFKL kernel are taken into
account, the situation changes conceptually due to the running of the QCD
coupling \cite{ck, cc2}. Running coupling effects mean that the saddle point
of $\chi$ is now a function of the scale $t=\log(k^2/\Lambda^2)$,
$\omega_s(t)=\bar{\alpha_s}(t)\chi_m$. Furthermore, $\omega_s(t)$ is not a
point of singularity of the Green's function $g_\omega(t, t_0)$, although the
value $\omega_s(t)$ does control the $Y$-dependence of $G$ over a limited
region of moderately large $Y$-values \cite{cc2}. The rightmost singularity of
$g_\omega(t, t_0)$ is at $\omega=\omega_P$, is independent of $t$ and $t_0$
and determines the asymptotically large $Y$-dependence of $G$, although for
$t$ and $t_0$ sufficiently large this asymptotic behaviour will not set in
until $Y$ is quite large. The singularity at $\omega_P$ is determined by
non-perturbative physics.

The fact that running coupling effects can change the character of the
$Y$-dependence of $G$ is easy to see. In the fixed coupling limit the BFKL
kernel gives a contribution proportional to $\alpha Y$ each time it acts. In
the running coupling case, the contribution is proportional to
$\alpha_s(t')Y\simeq \alpha_s(t)Y + b\log(t'/t)\alpha_s^2(t)Y$, when expressed
in terms of the fixed external scale $t$ of the scattering. However, the
contribution of a single running coupling term vanishes, because the average
value of $\log(t'/t)$ is zero since the probabilities for $t'>t$ and for $t<t$
are equal in fixed coupling BFKL evolution. At the level of two running
coupling contributions, one gets the average of $\alpha_s^4(t) Y^2
\log^2(t/t')\simeq \alpha_s^5 Y^3$ \cite{km,abb,ll,cc3}, and this contribution
exponentiates. This simple perturbative argument is valid so long as
$\alpha_s^5 Y^3 \ll 1$, but is difficult to extend to values of $Y >
t^{5/3}$. And it is exactly in the region $t^{5/3}< Y < t^2$ where the most
dramatic running coupling effects on BFKL evolution take place. 

In the present paper we calculate, starting from some small-$x$ models, the
diffusion and running coupling corrections to the hard-Pomeron
exponent. Basically, we prove the validity of the leading running coupling
corrections hinted at before \cite{km}--\cite{cc3}, in the full range $t\ll Y
\ll t^2$, and we discuss some features of the regime $Y \gappeq t^2$.

Because of the conceptual complexity caused by the running of the coupling, it
is useful to have a simple model where the essence of running coupling 
effects are 
present and yet a rather explicit discussion of the $Y$-dependence, in the
various regimes of $Y$, can be given. Such a model was introduced some time
ago \cite{cc2}, which takes into account the running of the coupling as well
as diffusion in momentum scales in terms of the quadratic dependence of
$\chi(\gamma)$ about the saddle point at $\gamma=1/2$ (Sec. 2). Here we study
the perturbative behaviour of this (Airy) model and of a simple generalization
where $\chi(\gamma)$ is represented as a sum of two poles in $\gamma$
\cite{cc3}. These two models give identical results for $G$ in the region $t
\ll Y \ll t^2$, and we believe they represent QCD accurately in this region, as
outlined below.

The solution of the Airy model is given in terms of Airy functions for the
perturbative part of the evolution and in terms of a reflection coefficient,
determined by the way the running coupling is regularized in the infrared, for
the non-perturbative part (Sec. 2). The singularity at $\omega_P$ resides in
the reflection coefficient $S(\omega)$ \cite{cc2}.

When $t \ll Y\ll t^{5/3}$ the behaviour of the perturbative part, $G_P(Y; t,
t_0)$, of $G$, is determined by a saddle point of the $\omega$-integral at
$\omega_s(t)$ giving $G_P\simeq \exp(\omega_s(t)Y)$. In this region of $Y$
running coupling effects play a minor role. When $Y$ reaches $t^{5/3}$ the
saddle point at $\omega_s(t)$ no longer gives the dominant contribution to
$G_P$. Nevertheless, the largest term in the exponent governing the
$Y$-dependence of $G_P$ is still $\omega_s((t+t_0)/2) Y$, with the next
largest corrections being given by the ``diffusion'' term and by the $Y^3/t^5$
term discussed above (Secs. 3, 4). This result is summarized by
Eq. (\ref{twentyone}) and by Eq. (\ref{fifty}) of the text, which give the
dominant behaviour throughout the region $t\ll Y\ll t^2$. When $Y$ reaches
$t^2$ the character of the solution changes drastically. For $Y\gg t^2$ the
magnitude of $G_P$ goes as $\exp(\bar{\omega}\sqrt{Y})$ where $\bar{\omega}$
is a pure, $t$-independent, number (Sec. 5). This behaviour comes from two
complex-valued saddle points having $\omega\sim 1/\sqrt{Y}$, whose exact
magnitude is model dependent (Sec. 6). Because the
saddle points are complex there is an oscillating prefactor in $G_P$, which
cannot be given a sensible physical interpretation, and calls for
non-perturbative contributions to take over.

The physical interpretation of our main results, Eqs. (\ref{twentyone}) and
(\ref{fifty}), seems clear. For $t\ll Y\ll t^{5/3}$, $G_P$ behaves - with 
good approximation -  as in the
fixed coupling case, having a magnitude proportional to $\exp(\omega_s(t)Y)$,
with a spread in $t-t_0 = \Delta t$ given by
$(\Delta t)^2 \sim \omega_s(t)Y$. At $Y\sim t^{5/3}$ running coupling effects
become more important. For $t$ fixed, $G_P(Y; t, t_0)$ reaches a maximum value
proportional to $\exp(\omega_s(t)+\eta^3/3)$ with $\eta \sim Y/t^{5/3}$,
while $\Delta t$ is well fixed at a value $\sim Y^2/t^3$, with only small
fluctuations of size $\sqrt{Y/t}$ from that value. Thus the values of $\Delta
 t$ are no longer given by pure diffusion, but now (for a fixed $t$) $t_0$ is
being pulled in the infrared where the coupling is large. When $Y$ gets as
large as $t^2$ the preferred value of $\Delta t$ becomes as large as $t$ and
the perturbative part of the model ceases to make physical sense.

Finally, a comment on the accuracy to which we have calculated $G_P$. It is
convenient to write $\log{G_P}=\omega_sY~f(Y^2/t^4,\Delta t /t)+$ small
terms. We have calculated terms in $f$ of size $1$, $Y^2/t^4$, $\Delta t/t$, 
and
$(\Delta t)^2~t^2/Y^2$, as given in Eq. \ref{twentyone}. In the dominant
region of $\Delta t \sim Y^2/t^3$ all these corrections are of size
$Y^2/t^4$. We certainly expect further corrections to $f$ of size $Y^4/t^8$,
as discussed in Secs. 3 and 4. So long as $ Y\ll t^2 $ we believe we have the
dominant terms in the exponent and so have the essence of the growth of $G_P$
with $Y$ and its dependence on $\Delta t$. However, we only have some
preliminary ideas on how to calculate corrections beyond this region (Sec. 5),
and thus we likely do not have all the large terms in the exponent of $G_P$. 
 
\section{The gluon density in small-$x$ models}

We consider in this paper a particular form of the 2-scale gluon Green's
function which has been established for the Airy model \cite{cc2} and for
the
truncated BFKL models \cite{cc3}. By defining
\beq
G(Y; k, k_0) = {1\over kk_0} \int {d\omega\over 2\pi i} e^{\omega Y} g_\omega
(t,
t_0)~,
\label{one}
\eeq
where $Y \equiv \log {s\over kk_0}$ and $t_i \equiv \log {k^2_i\over
\Lambda^2}$, $g_\omega$ takes a factorized form for $t > t_0$, namely
\beq
g_\omega(t, t_0) = {\cal F}_\omega(t) \bigg(\tilde{\cal F}_\omega(t_0) +
S(\omega ) {\cal F}_\omega (t_0)\bigg)~, ~~~ (t > t_0)~,
\label{two}
\eeq
where the various terms are defined as follows:\\
- ${\cal F}_\omega(t)$ is the ``regular" solution of the homogeneous
small-$x$
equation being considered, which vanishes exponentially for large $t$
values;
\\
- $\tilde{\cal F}_\omega(t_0)$ is an ``irregular" solution, which is instead
exponentially increasing with $t_0$;\\
- $S(\omega )$ is a ``reflection" coefficient, which has been explicitly
constructed in some cases \cite{cc2}, \cite{cc3} and is dependent on the 
strong coupling
region, e.g., on how the $t_0=0$ Landau pole is smoothed out or cutoff.

While the explicit form and size of the non-perturbative $S(\omega )$ part
is
dependent on the model -- in particular on the number of poles taken into
account in the effective eigenvalue function \cite{cc3}, the perturbative
term
is unambiguously defined in the large-$t$ region, and is supposed to be
calculable in a realistic small-$x$ framework \cite{ccs}. For this reason,
most
of our analysis will concern the perturbative term.

In order to understand better the meaning of Eq. (\ref{two}), let us derive
it explicitly in the Airy model. The defining equation for the
Green's function is, in operator notation,
\beq
(\omega - \bar\alpha_s K) g_\omega = 1
\label{three}
\eeq
where $K$ is in general an integral kernel, and $\bar\alpha_s(t)
\equiv N_C \alpha_s(t)/\pi \equiv 1/bt$ is the running coupling.
The Airy model obtains by assuming that $K$ is scale invariant, and
described
by a quadratic eigenvalue function
\beq
\chi (\omega ) \simeq \chi_m + {1\over 2} \chi_m''  (\gamma - {1\over
2})^2~,
~~~ (\chi_m, \chi_m'' > 0)~,
\label{four}
\eeq
which is a reasonable approximation around the minimum of $\chi$, which is
taken to be at $\gamma = {1\over 2}$. Here $\gamma$ is a variable conjugated
to $t$ by Fourier transform. Therefore, by using Eq. (\ref{four})
in $t$-space, Eq. (\ref{three}) becomes
\beq
\bigg[\omega - \omega_s(t) ~(1+D\partial^2_t)\bigg] g_\omega(t,t_0) = \delta
(t - t_0)~,
\label{five}
\eeq
where $\omega_s(t) = \chi_m \bar\alpha_s(t)$ is the hard Pomeron
exponent, and $D = {1\over 2} \chi_m''/\chi_m$ is the diffusion coefficient.

In other words, the Airy Green's function satisfies a second order differential
equation in the $t$ variable and has, therefore, the well-known form
\beq
g_\omega(t,t_0) = t_0 \bigg({\cal F}^R_\omega(t) {\cal
F}^L_\omega(t_0)~~\Theta(t-t_0) + (t \leftrightarrow t_0)\bigg)
\label{six}
\eeq
where ${\cal F}^R_\omega(t) (t_0{\cal F}^L_\omega(t_0))$ is the regular 
solution of
the homogeneous equation in (\ref{five}) for $t\rightarrow\infty$ (of the
adjoint homogeneous equation for $t_0\rightarrow -\infty)$. Equation
(\ref{six}) is the basis for Eq. (\ref{two}), but should be better specified
by smoothing out or cutting off $\bar\alpha_s(t)$ in a region $t\lappeq \bar
t$, where $\bar t > 0$ defines the boundary of the perturbative regime
$\bar\alpha_s(t) = 1/bt$. Depending on such procedure, the left-regular
solution can be evaluated for large $t_0$ values in the form
\beq
{\cal F}^L_\omega(t_0) = {\cal F}^I_\omega(t_0) + S(\omega) {\cal
F}^R_\omega(t_0)~,
\label{seven}
\eeq
where ${\cal F}^I$ is irregular for $t_0 \rightarrow \infty$, and
$S(\omega)$ is a well-defined reflection coefficient. We have thus
derived Eq. (\ref{two}) with ${\cal F} = {\cal F}^R$ and $\tilde{\cal F} =
t_0{\cal F}^I$. A similar derivation holds for the truncated BFKL models
with
$n$ poles, and in particular for the 2-pole model (Sec. 6).

In the Airy model, Eqs. (\ref{six}) and (\ref{seven}) can be given in
explicit
form. In the perturbative region $t > \bar t$ we can set, by Eq.
(\ref{five}),
\beq
{\cal F}_\omega(t) = {\cal F}^R_\omega(t) = Ai(\xi) =
\int^{+i\infty}_{-i\infty}~~{d\nu\over 2\pi i}~~e^{\nu\xi - {1\over
3}\nu^3} = {\sqrt{\xi}\over \pi\sqrt{3}}~~K_{1/3} \bigg({2\over 3}
\xi^{{3\over 2}}\bigg)~,
\label{eight}
\eeq
and furthermore,
\beq
\tilde{\cal F}_\omega(t_0) = t_0 {\cal F}^I_\omega(t_0) = {\pi
~t_0\over\omega}~\bigg( {2b\omega\over \chi^"_m}\bigg)^{{2\over 3}} Bi
(\xi_0)~,~~~S(\omega) = -{Bi(\bar\xi)\over Ai(\bar\xi)}~,
\label{nine}
\eeq
where $Ai(\xi)$ is the Airy function,
\beq
Bi(\xi) = e^{{i\pi\over 6}} Ai(\xi e^{{2i\pi\over 3}} ) + e^{{-i\pi\over 6}}
Ai(\xi e^{{-2i\pi\over 3}})
\label{ten}
\eeq
is the irregular Airy solution,
\beq
\xi= \bigg({2b\omega\over \chi_m''}\bigg)^{1\over 3}~ \bigg( t - {\chi_
m\over b\omega}\bigg) = D^{-{1\over 3}} t^{2\over 3} \bigg(
{\omega_s(t)\over
\omega}\bigg)^{{2\over 3}}~ \bigg({\omega\over\omega_s(t)} -1\bigg)
\label{eleven}
\eeq
is the Airy variable, and $\bar\xi$ is its value for $t = \bar t$, at the
boundary of the perturbative region. Note that, given the delta-function
source in (\ref{five}), the regular and irregular solutions in Eqs.
(\ref{eight}) and (\ref{nine}) must have a well defined Wronskian.

From the explicit expressions (\ref{eight})-(\ref{ten}) it follows that
${\cal
F}_\omega$ and  $\tilde{\cal F}_\omega$ are analytic functions of $\omega$,
showing an essential singularity at $\omega = 0$ only. Instead, the
reflection coefficient $S(\omega)$ -- quoted in Eq. (\ref{nine}) in the case
$\alpha_s(t)$ is cutoff at $t = \bar t$ -- shows $\omega$ singularities in
Re$\omega > 0$ at the zeros of the Airy function, meaning that the leading
Pomeron singularity $\omega_P$ actually occurs in the non-perturbative term
in
Eq. (\ref{two}). However, the latter is suppressesd, at large $t_0$, by the
ratio $Ai(\xi_0)/Bi(\xi_0)$, meaning that the perturbative term may actually
be more important at large scales and intermediate energies.

By using the decomposition in Eq. (\ref{two}) we can rewrite Eq. (\ref{one})
as a sum of two terms
\beq
G(Y ; k, k_0) = G_P + G_R~,
\label{twelve}
\eeq
where
\beq
G_R(Y;k,k_0) = {1\over kk_0} \int_{C_R} {d\omega\over 2\pi i} e^{\omega Y}
S(\omega)~ {\cal F}_\omega (t) {\cal F}_\omega t_0)
\label{thirteen}
\eeq
carries the (power behaved) Regge contributions (Fig. 1), the leading one
being
the (non-perturbative) Pomeron, while
\bea
G_P(Y;k,k_0) &=& {1\over kk_0} \int_{C_s} {d\omega \over 2\pi i} e^{\omega Y} 
{\cal F}_\omega(t) \tilde{\cal F}_\omega (t_0) \nonumber \\
&\simeq& {1\over kk_0} \exp (\omega_s(kk_0) Y + \ldots )
\label{fourteen}
\eea
corresponds to the ``background integral" and is characterized by the
two-scale exponent $\omega_s$ that will turn out to be determined by
$\alpha_s(kk_0)$ (Section 3).
%\newpage
\begin{figure}
%\vspace*{-1cm}
\centerline{\psfig{figure=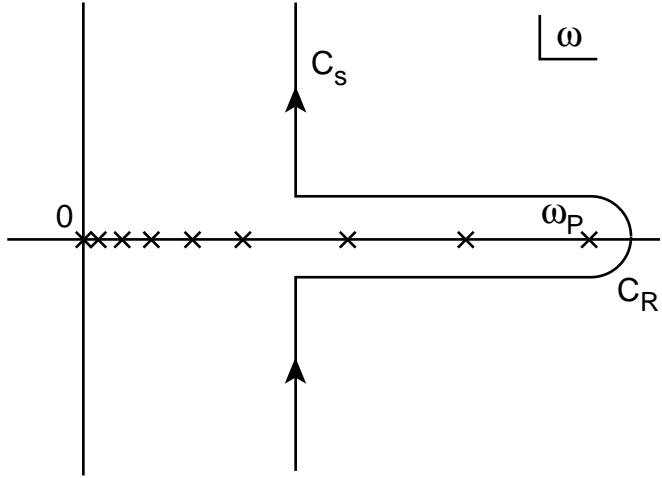}}
%\vspace*{-1cm}
\caption{Integration contour $C_R+C_s$ for the Green's function decomposition
  in Eq. (\ref{twelve}). $C_R(C_s)$ refers to the Pomeron (perturbative)
  contribution.} 
\end{figure}
%\newpage
Our goal in the following is to analyze in more detail the $Y$-dependence in
Eq. (\ref{fourteen}), by determining the regime in which the exponent
$\omega_s$ is relevant, and the magnitude and form of diffusion corrections
to it.

\section{Diffusion features of the Airy model: a heuristic \\approach}

We have just clarified that the two terms in the decomposition (\ref{two})
generate the $Y$-dependence in $G(Y; k, k_0)$ in a
quite different way. The non-perturbative (Pomeron) term provides just a
Regge-pole behaviour $\sim \exp (\omega_PY)$, while the perturbative one is
analogous to a ``background integral" (Fig. 1) and will generate a
non-trivial
exponent only if the small-$\omega$ oscillations of ${\cal F}$ and
$\tilde{\cal F}$ are kept in phase by the $\omega$-integration. By writing, 
for the Airy model,
\beq
g_P(Y;t,t_0) \equiv kk_0~G_P = \int {d\omega\over 2\pi i}~e^{\omega Y} {\pi
t_0\over \omega}~~\bigg({2b\omega\over \chi^"_m}\bigg)^{2\over 3}
Ai(\xi)~Bi(\xi_0)~,
\label{fifteen}
\eeq
we expect, for $\Delta t = t - t_0 \ll t$ that phase relations are kept only
if $\xi$ and $\xi_0$ are kept finite for large $Y$. By the definition
(\ref{eleven}), this implies that $\omega - \omega_s(t) \simeq \omega -
\omega_s(t_0) = 0(t^{-{5\over 3}})$ are small parameters. Furthermore, in
this
region, by Eq. (\ref{eleven}),
\beq
\xi \simeq D^{-{1\over 3}} t^{{2\over 3}}~~\bigg({\omega\over\omega_s(t)}
-1\bigg)
\label{sixteen}
\eeq
is just linear in $\omega - \omega_s$. By replacing the linearized
expression
(\ref{sixteen}) into Eq. (\ref{fifteen}), we can rewrite it in the form
\beq
g_P(Y;t,t_0) \simeq \pi (Dt)^{-{1\over 3}}
e^{\omega_s(t)Y}~~\int^{+i\infty}_{-i\infty}~{d\xi\over 2\pi i} ~ e^{\xi\eta} 
Ai(\xi)Bi (\xi-\delta)~,
\label{seventeen}
\eeq
where we have introduced the parameters
\beq
\eta = D^{{1\over 3}}~Y~\omega_s(t)~t^{-{2\over 3}} \sim Y t^{-{5\over 3}}~,
~~~\delta = \Delta t(Dt)^{-{1\over 3}}~.
\label{eighteen}
\eeq
\begin{figure}
\centerline{\psfig{figure=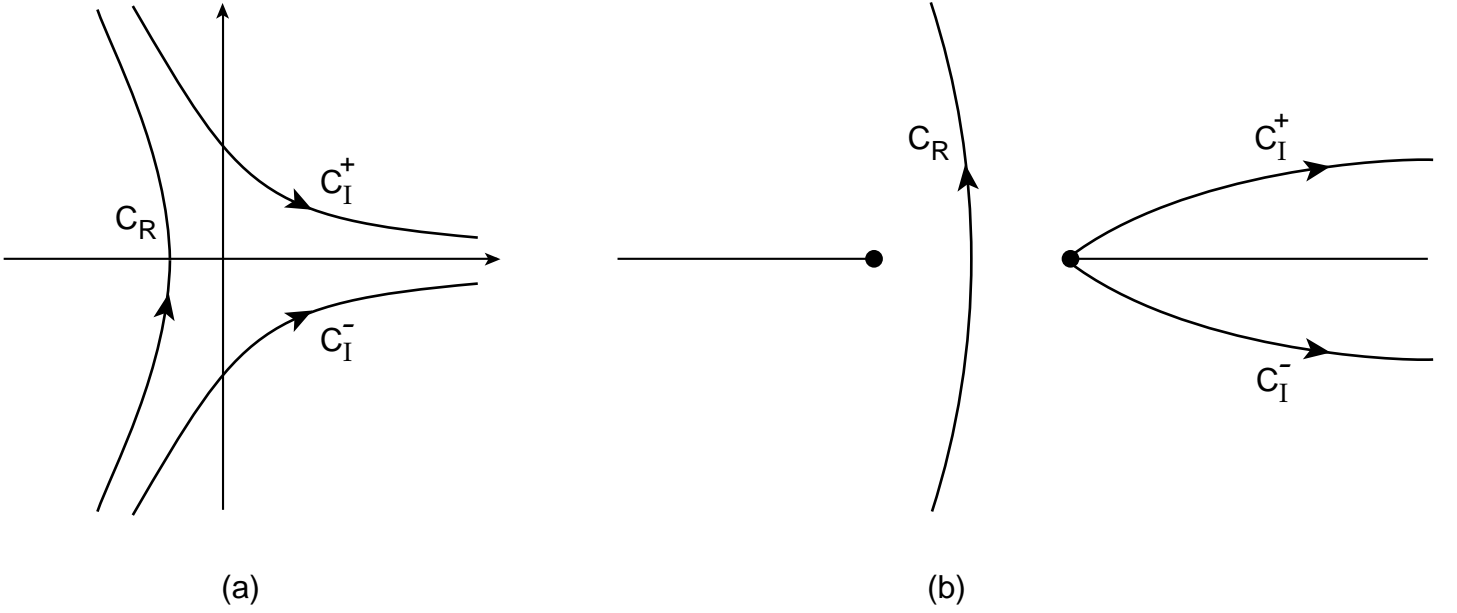}}
\caption{Contours in the $\nu=\gamma-{1/2}$ plane for ($C_R$) the regular and
  ($C_I$) the irregular solution in the case of ($a$) the Airy model and ($b$)
  the two-pole collinear model.}
\end{figure}

The integral in (\ref{seventeen}) can now be evaluated in a heuristic way by
introducing the integral representations (Fig. 2a)
\beq
Ai(\xi) = \int_{C_R}~{d\nu_A\over 2\pi i}~e^{\nu_A\xi - {1\over 3}
\nu_A^3}~,
Bi(\xi) = \int_{C_I^++C^-_I}~{d\nu_B\over 2\pi }~e^{\nu_B\xi - {1\over 3}
\nu_B^3}~
\label{nineteen}
\eeq
and by performing the $\xi$-integration under the integral, to yield a
delta-function. We thus obtain
\bea
I(\eta,\delta) \equiv \int~{d\xi\over2\pi
i}~e^{\xi\eta}~Ai(\xi)~Bi(\xi-\delta) &=&
\int^{+i\infty}_{-i\infty}~{d\nu_A\over 2\pi^2 i}~\exp \bigg(\eta \nu^2_A +
(\eta^2+\delta) \nu_A + \eta\delta + {\eta^3\over 3} \bigg)
\nonumber \\
&=& {1\over 2\pi \sqrt{\pi\eta}}~\exp \bigg( {\eta^3\over 12} +
{\eta\delta\over 2} - {\delta^2\over 4\eta}\bigg)~.
\label{twenty}
\eea
By inserting Eqs. (\ref{twenty}) and (\ref{eighteen}) into Eq.
(\ref{seventeen}), we finally get
\beq
g_P(Y,t,t_0) = {1\over \sqrt{4\pi D\omega_s(t)Y}}~\exp \bigg[ \omega_s(t) Y
\bigg( 1+{\Delta t \over 2t}\bigg) - {(\Delta t)^2\over 4D \omega_s(t)Y} +
{\eta^3\over 12}\bigg]
\label{twentyone}
\eeq
which provides the diffusion and running coupling corrections to the hard 
Pomeron behaviour.

There are two features worth noting in Eq. (\ref{twentyone}), in comparison
with the customary expression with frozen coupling. Firstly, the
exponent $\omega_s$ is corrected by a term linear in $\Delta t = t - t_0$
which provides the symmetrical argument $\omega_s({t+t_0\over 2})$ in the
running coupling, as is appropriate for the factorized scale $\sim k k_0$
already introduced in eq. (\ref{twelve}). Furthermore, the exponent carries
the diffusion correction $\eta^3 \sim Y^3/t^5$, which is of relative order
$Y^2/t^4$ compared to the leading term $\omega_s(t)Y$. This correction was
obtained as running coupling effect in \cite{km}, \cite{abb},\cite{ll}and 
confirmed \cite{cc3} in
the models considered here under the assumption $\eta\ll 1$, or $Y \ll
t^{{5\over 3}}$.

The question then arises of the boundary of validity of the heuristic
argument
presented above. We shall see in Section 4 that the assumption $\eta\ll 1$
can be relaxed, and replaced by $\eta\ll t^{{1\over 3}}$, or $Y \ll t^2$.
For
$Y \gappeq t^2$ on the other hand, the linearization in Eq. (\ref{fourteen})
breaks down, and the integral in Eq. (\ref{thirteen})
enters a new regime in which it first decreases, and then starts oscillating
(Section 5), so that the phase relations are lost.

A first hint at such behaviour is obtained by replacing the ansatz in Eq.
(\ref{twentyone}) in the diffusion equation
\beq
{\partial\over\partial Y}~g_P (Y,t,t_0) \simeq \omega_s(t)~\bigg(1+D
\partial^2_t\bigg) ~ g_P(Y,t,t_0)~.
\label{twentytwo}
\eeq
In fact, the $Y$-derivative of the exponent $E(Y,t,\Delta t) = 
\log g_P$ is given by
\beq
{\partial E(Y,t, \Delta t)\over \partial Y} = -{1\over 2Y} +
\omega_s(t)~\bigg(1+{\Delta
t\over 2t}\bigg) + {(\Delta t)^2\over 4 D\omega_s Y^2} +
{\partial\over\partial Y}~\bigg({\eta^3\over 12}\bigg)
\label{twentythree}
\eeq
and should match the right-hand side of Eq. (\ref{twentytwo}), which is
given
by
\bea
&&\omega_s(t) \bigg[ 1 + D\bigg(({\partial\over \partial t}E)^2 +
{\partial^2E\over \partial t^2}\bigg)\bigg]~,\nonumber \\
&&{\partial E\over \partial t} = -{1\over 2t} \omega_s(t)Y(1+{\Delta t\over
t}) - {\Delta t\over 2D\omega_sY}~(1+2{\Delta t \over t}) +
{\partial\over\partial t}~({\eta^3\over 12}) + {1\over 2t}~.
\label{twentyfour}
\eea

If we keep terms of relative order $t/Y$, $\Delta t/t$, $Y^2/t^4$ (such
terms
are all of the same order for fixed values of $\eta$), we find
that Eqs. (\ref{twentythree}) and (\ref{twentyfour}) are indeed consistent,
provided
\beq
{\partial\over\partial Y}~~{\eta^3\over 12} = D {\omega^3_s Y^2\over 4t^2}~,
\label{twentyfive}
\eeq
thus reproducing the expression (\ref{eighteen}) for $\eta(Y,t)$. But,
because
of (\ref{twentyfive}), we also generate from Eq. (\ref{twentytwo}) terms of
relative order $Y^4/t^8$ which are subleading only if $Y \ll t^2$, and do
not
check with Eq. (\ref{twentythree}). Therefore, the heuristic argument breaks
down for $Y = 0(t^2)$.

\section{Detailed analysis of the regime $t < Y \ll t^2$}

In this section we give a more complete evaluation of the integral
(\ref{fifteen}) which defines $g_P$ and in doing so we confirm the result
(\ref{twentyone}) and its validity boundary, obtained in a heuristic way in
the previous section.

\subsection{Choice of integration contour}
We begin with the $\omega$-integration contour in (\ref{thirteen}) being
parallel to the imaginary $\omega$-axis with Re$\omega = \omega_s$ (Fig. 3). 
In order
to effectively separate leading from non-leading behaviours in
(\ref{thirteen})
it is convenient to use different forms of the product $Ai(\xi)~Bi(\xi_0)$
for
Im$ \omega > 0$ and for Im$ \omega < 0$. To that end we write
\beq
Ai(\xi)Bi(\xi_0) = \mp {2i\sqrt{\xi\xi_0}\over 3\pi^2}~K_{{1\over 3}}
({2\over 3} \xi^{{3\over 2}})~\bigg[ K_{{1\over 3}}({2\over 3}\xi_0^
{{3\over 2}}e^{\mp
i\pi}) -{1\over 2}~ K_{{1\over 3}} ({2\over 3} \xi_0^{{3\over 2}})\bigg]
\label{twentysix}
\eeq
where the upper (lower) sign will be the form used in (\ref{thirteen}) when
Im$ \omega > 0 (<0)$. Equation (\ref{twentysix}) follows from (\ref{eight})
and (\ref{ten}) along with
\beq
K_\nu(e^{i\pi}\zeta)+K_\nu(e^{-i\pi}\zeta) = 2 \cos \pi \nu K_\nu(\zeta)~.
\label{twentyseven}
\eeq
Thus
\beq
g_P(Y_;t,t_0) = \bar g_P(Y;t.t_0) + R(Y;t,t_0)
\label{twentyeight}
\eeq
with
\beq
\bar g_P(Y;t,t_0) = \int \mp {t_0d\omega\over 3\pi^2\omega
D}~\sqrt{({\omega\over\omega_s(t)}-1)~({\omega\over\omega_s(t_0)}-1)}
~~e^{\omega Y} K_{{1\over 3}} ({2\over 3} \xi^{{3\over 2}})
 ~K_{1\over 3}({2\over 3} \xi_0^{{3\over 2}}
e^{\mp i \pi})
\label{twentynine}
\eeq
and
\beq
R(Y;t,t_0) = \int \pm {t_0d\omega\over 6\pi^2\omega
D}~\sqrt{({\omega\over\omega_s(t)}-1)~({\omega\over\omega_s(t_0)}-1)}
~~e^{\omega Y} K_{{1\over 3}} ({2\over 3} \xi^{{3\over 2}})
 ~K_{1\over 3}({2\over 3} \xi_0^{{3\over 2}})
\label{thirty}
\eeq
where the upper (lower) sign is to be used when Im$\omega > 0(<0)$.

We shall first show that $R$ can be chosen small compared to $\bar g_0$ by a
judicious contour deformation. Since the integrand of (\ref{twentyfour})
decreases for positive real $\xi$, we are led to deform $C_s$ around the
real $\xi$ axis to reach a point $\xi_1>1$ to be defined below.
In estimating the size of $R$ we shall take $\xi_0 \approx \xi$ so that the
full exponent appearing in (\ref{thirty}), in the region where $|\xi|\gg 1$
where the asymptotic form of $K_{{1\over 3}}$ can be used, is
\beq
{\cal E} = \omega_s(t) Y + (\omega - \omega_s(t)Y - {4\over 3} \xi^{{3\over
2}}~.
\label{thirtyone}
\eeq
In the region where $\xi \ll t^{{2\over 3}}$
\beq
{\cal E} - \omega_s Y \simeq \xi\eta - {4\over 3} \xi^{{3\over 2}}~.
\label{thirtytwo}
\eeq
Thus we can make the second term on the right-hand side of (\ref{thirtytwo})
dominate the first term at $\xi = \xi_1$ if
\beq
\xi_1 / \eta^2 \gg 1~.
\label{thirtythree}
\eeq
We note that (\ref{thirtythree}) is possible, keeping $\xi_1 \ll t^{2/3}$, 
so long as $\eta\ll t^{{1\over 3}}$ or, equivalently, so long as $Y \ll
t^2$. We anticipate $\bar g_P$ being of size $e^{\omega_sY}$ so that if we
are
able to choose an integration path in (\ref{twentyfour}) having Re${4\over
3} \xi^{{3\over 2}}\gg (\omega - \omega_s)Y$, then $R$ can be neglected.
\begin{figure}
\centerline{\psfig{figure=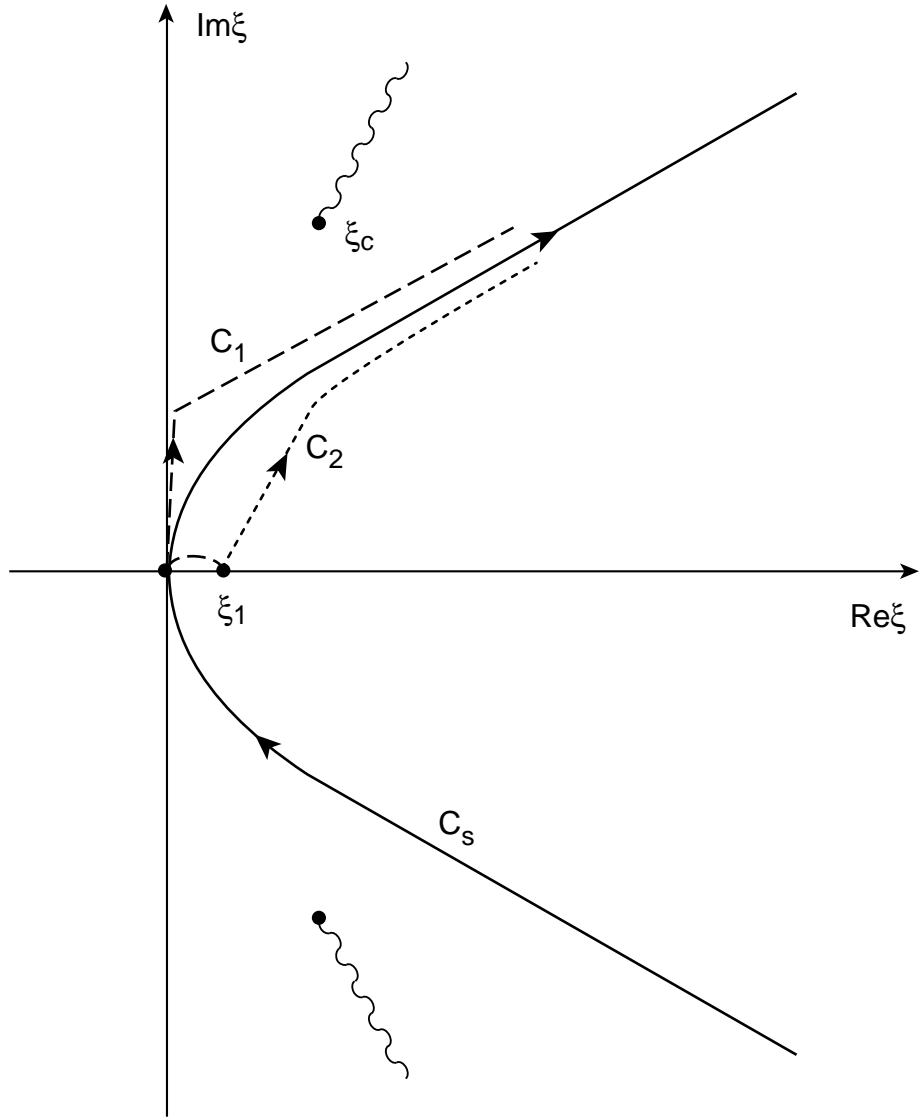}}
\caption{Integration contours in the $\xi$ plane, cut at $\xi_c\sim
  t^{2/3}$. $C_s$ corresponds to Re$\omega=\omega_s$. The dashed (dotted)
  contour $C_1$ ($C_2$) corresponds to the deformed contour defined in the
  text. The corresponding ones in Im$\xi<0$ are not shown}
\end{figure}
A contour of this kind is $C_2$ in Fig. 3. It is basically a deformation of the
contour $C_s$ at Re$\omega = \omega_s$ in order to have $\xi_1$ as starting
point and to depart from it with Arg$(\xi - \xi_1) < {\pi\over 3}$. Its
basic
property is to lie completely within the regions $|\rm Arg \xi| 
< {\pi\over 3}$,
such that the product of regular Airy functions is damped (Re$\xi^{{3\over
2}} > 0)$. It is then easy to convince oneself that Re$({4\over 3}
\xi^{{3\over 2}})\gg$ Re$(\omega - \omega_s)Y$ on a contour of this kind,
so
that the contribution $R$ in Eq (\ref{thirty}) can be neglected compared to
$\bar g_{P}$ so long as $Y \ll t^2$.

\subsection{From the $\omega$-integral to the $\xi$-integral}

Our task is now to evaluate (\ref{twentynine}) and to specify once again a
convenient path of integration in $\xi$, that will turn out to be different
from that chosen before. We are able to evaluate (\ref{twentynine}) only in
a
linear approximation of $\xi$ and $\omega - \omega_s(t)$, and we first turn
to
a more complete justification of this approximation. When $|\xi_0|$ and
$|\xi|$ are large, the terms appearing in the exponent in (\ref{twentynine})
are
\beq
{\cal E} = \omega Y - {2\over 3} \xi^{{3\over 2}} + {2\over 3} \xi_0^{{3\over
2}}~.
\label{thirtyfour}
\eeq
This can be written as
\beq
{\cal E} = \omega_s(t) Y + \xi\eta\bigg({\omega\over\omega_s}\bigg)^{{2\over
3}} - {2\over 3}~\xi^{{3\over 2}} + {2\over 3} (\xi - \delta
\bigg({\omega\over\omega_s}\bigg)^{{1\over 3}})^{{3\over 2}}~,
\label{thirtyfive}
\eeq
with $\delta$ as given in (\ref{eighteen}). The linear approximation gives
\beq
{\cal E}_{\rm lin} = \omega_s(t) Y + \xi\eta - {2\over 3}~\xi^{{3\over 2}} +
{2\over
3} (\xi - \delta)^{{3\over 2}}~,
\label{thirtysix}
\eeq
In order to see how close ${\cal E}$ and ${\cal E}_{\rm lin}$ are, we deform
$C_s$ to the contour $C_1$, in this case (Fig. 3). The latter, departing
from
$\xi_1$ and reaching the origin, is chosen in such a way as to run over the
imaginary $\xi$ axis in the whole linearization region $|\rm Im \xi| < 
t^{{2\over3}}$, and to approach $C_s$ later on in the Re$\omega < \omega_s$  
region.
It is straightforward to see that the regions of $C_1$ on the real and
imaginary axis are the dominant ones in the integral. In fact, from Eq.
(\ref{thirtysix}) one sees that Re$({\cal E}_{\rm lin} -\omega_sY)$ vanishes at
$\xi = 0$ and decreases steadily below zero for $\xi$ on the imaginary axis.
Furthermore, in the asymptotic region where $C_1$ starts approaching $C_s$
the
Airy phase is negligible and the integrand in Eq. (\ref{twentynine}) is
damped
because of Re$\omega < \omega_s$.

Now we want to replace ${\cal E}$ by ${\cal E}_{\rm lin}$ in the exponent. We
can
estimate the error in the exponent by comparing (\ref{thirtyfive}) and
(\ref{thirtysix}) in the region $0 \leq \xi \leq \xi_1$. The maximum
deviation
occurs for $\xi = \xi_1$ and the size of the deviation is
\beq
\Delta{\cal E} = {\cal E} - {\cal E}_{\rm lim} \simeq \xi_1\eta
\bigg[\bigg({\omega\over\omega_s}\bigg)^{{2\over 3}} -1\bigg] -
\sqrt{\xi_1}\delta \bigg[\bigg({\omega\over\omega_s}\bigg)^{{1\over 3}} -1
\bigg]~.
\label{thirtyseven}
\eeq
From (\ref{eleven}) one sees
$\bigg[\bigg({\omega\over\omega_s}\bigg)^{{1\over
3}} -1\bigg]$ and $\bigg[\bigg({\omega\over\omega_s}\bigg)^{{2\over 3}} -1
\bigg]$ are of size $\xi_1 t^{-2/3}$, so that $\Delta {\cal E}$ has 
terms of size $\xi_1^2 \eta t^{-2/3}$ and
$\xi^{3/2}_1 {\Delta t /t}$. Taking $\xi_1 =
N\eta^2$ for $N$ large, the smallest we are allowed to take $\xi_1$, and
taking $\Delta t \simeq (Dt)^{{1\over 3}} \eta^2$ as region of $\Delta t$
where $g_P$, as given in (\ref{nineteen}), take its maximum value we find
\beq
\Delta{\cal E} \approx {\eta^5\over t^{{2\over 3}}}~~,~~\sim \omega_sY 
{Y^4\over t^8}.
\label{thiertyeight}
\eeq
Thus we can expect our final result and have errors of size $\Delta{\cal E}$
in the exponent.

In changing the exponent ${\cal E}$ in (\ref{twentynine}) to ${\cal E}_{\rm
lin}$ we take the same contour as indicated in Fig. 3, but now the
integration
is continued along the imaginary $\xi$-axis up to $\xi = i\infty$. Using the
linearized form for $d \xi$ also
induces an error in the prefactor of the dominant exponential of size
${\omega
- \omega_s\over \omega_s}\propto \bigg({\eta\over t^{{1\over 3}}}\bigg)^2
\sim
{Y^2\over t^4}$ compared to 1. Thus, finally, the integration which we need
to evaluate is
\beq
g_P(Y;t,t_0) = \mp [Dt]^{-{1\over 3}} e^{\pm i{\pi\over 3}}
e^{\omega_s(t)Y}~~\int_{C_1} d\xi~ e^{\eta\xi}~Ai(\xi)
Ai\bigg((\xi-\delta)e^{\mp 2i{\pi\over 3}}\bigg)
\label{thirtynine}
\eeq
where we have used (\ref{eight}) to express the $K_{{1\over 3}}$ functions
appearing in (\ref{twentythree}) by Airy functions for reasons which will
become clear in a moment. The integration is taken with the contour $C_1$
extended to $\pm i\infty$, with, as usual, the upper (lower) sign referring
to
Im$\xi$ greater (less) than zero.

\subsection{Evaluation of the $\xi$-integrals}

In order to evaluate the integral in (\ref{thirtynine}) it is convenient to
define
\beq
I_\pm = \int^{\pm i\infty}_{\xi_1\pm i\epsilon} d\xi~e^{\eta\xi} P_\pm (\xi)~,
~~~
J_\pm = \int^{\pm i\infty}_{\xi_1\pm i\epsilon} d\xi~e^{\eta\xi} Q_\pm (\xi)~,
\label{forty}
\eeq
\beq
P_\pm(\xi) = U(\xi) V_\pm (\xi-\delta)~,~~~Q_\pm (\xi) =
{\partial\over\partial \xi}U(\xi) V_\pm(\xi-\delta)~,
\label{fortyone}
\eeq
with

\beq
U(\xi) = Ai(\xi)~,~~~ V_\pm(\xi) = \mp e^{\pm i{\pi\over
3}}~Ai((\xi-\delta)~e^{\mp 2i{\pi\over 3}})~.
\label{fortytwo}
\eeq
Using
${d^2\over d\xi^2} U(\xi) \equiv U''(\xi) = \xi U(\xi)$ and
$V_\pm'' (\xi-\delta) = (\xi-\delta) V_\pm (\xi-\delta)$ one can show that
\beq
P''' = [4 \xi P - \delta P]^{'} -2P -2\delta Q~,~~~ Q'' = 2(\xi P)^{'} -P
-\delta Q
\label{fortythree}
\eeq
from which, by integration by parts, one finds
\bea
2I &=&- \eta [4{\partial\over\partial\eta} - \delta - \eta^2]~I -2\delta J +
e^{\eta\xi_1} P_2(\xi_1,\eta)~, \nonumber \\
J(\delta + \eta^2) + I &=& -2\eta {\partial I \over \partial\eta} +
e^{\eta\xi_1} Q_2(\xi_1\eta)~,
\label{fortyfour}
\eea
where the $\pm$ indices have been suppressed in
(\ref{fortythree})-(45). Also
\bea
P_2 &=& P''(\xi_1) - \eta P^{'}(\xi_1) + \eta^2P(\xi_1) - (4\xi_1 - \delta)
P(\xi_1)~,
\nonumber \\
Q_2 &=& Q^{'}(\xi_1) -2\xi_1 P(\xi_1) -\eta Q (\xi_1)~.
\label{fortyfive}
\eea
Using Eq. (\ref{fortyfour}), one easily finds
\beq
\bigg[ \eta{\partial\over\partial\eta} + {1\over 2} - {(\delta +
\eta^2)^2\over
4\eta}\bigg] ~I_\pm = {e^{\eta\xi_1}\over 4\eta^2}
\bigg[(\delta+\eta^2)P_{2\pm} -
2\delta Q_{2\pm}\bigg] = S_\pm
\label{fortysix}
\eeq
The solution to (\ref{fortysix}) is
\beq
I_\pm = {1\over\sqrt{\eta}}~\exp \bigg[{\eta^3\over 3} -
{(\delta-\eta^2)^2\over 4\eta}\bigg]~\bigg(C_\pm +
\int^\eta_{\eta_0}~{d\eta^{'}\over\sqrt{\eta^{'}}}~\exp \bigg[
-{\eta^{'3}\over 3}
+ {(\delta-\eta^{'2})^2\over 4\eta^{'}}\bigg] S_\pm (\eta^{'}\bigg)
\label{fortyseven}
\eeq
with $C_\pm$ a constant to be determined. Referring back to
(\ref{thirtynine})
one can write
\beq
g_P (Y;t,t_0) = (D^2 t)^{-{1\over 3}}~e^{\omega_s(t)Y}~(I_+-I_-)~.
\label{fortyeight}
\eeq

Now the combination $S_+-S_-$ occurring in the integrand of the particular
solution of (\ref{fortyseven}) which enters in (\ref{fortyeight}) always
involves the combination
$$
K_{1\over 3}({2\over 3}(\xi_1-\delta)^{3\over 2} ~e^{-i\pi}) +
K_{1\over 3}({2\over 3}(\xi_1-\delta)^{3\over 2} ~e^{+i\pi}) =
K_{1\over 3}
({2\over 3}(\xi_1-\delta)^{3\over 2}) \propto \exp
[-{2\over 3} (\xi_1 - \delta)^{3\over 2} ]
$$
so that the particular solution can be ignored. Although we have focused
here
on $\eta$ large the procedure also should be valid when $\eta \ll 1$. For
$\eta
\ll 1$, the first factor in (\ref{fortyseven}),
${1\over\sqrt{\eta}}\exp[-{\delta^2\over 4\eta}]$, gives the usual BFKL
diffusion which means that $C_+-C_-$ can be determined by evaluating
$I_+-I_-$
when $\delta = 0$ and when $\eta$ is small. One finds
\beq
C_+-C_- = {1\over \sqrt{4\pi}}
\label{fortynine}
\eeq
giving exactly the result (\ref{twentyone})
\beq
g_P(Y;t,t_0) ~=~ {1\over \sqrt{4\pi D\omega_s(t)Y}}~e^{\omega_s(t)Y-
{1\over 4\eta}(\delta -\eta^2)^2+{\eta^3\over 3}}
\label{fifty}
\eeq

In arriving at this result we have chosen to expand the integrand of
(\ref{twentynine}) about $\xi$ and $t$ and then integrate over $\xi$.
Alternatively, if the expansion had been about $\xi_0$ and $t_o$ with the
integration over $\xi_0$ the result
\beq
g_P(Y;t,t_0) = {1\over \sqrt{4\pi
D\omega_s(t_0)Y}}~e^{\omega_s(t_0)Y-{1\over
4\eta_0}
  (\delta_0 -\eta^2_0)^2+{\eta^3_0\over 3}}
\label{fiftyone}
\eeq
would have been obtained with $\eta_0 = Y \omega_s(t_0) D^{{1\over 3}}
t_0^{-{2\over 3}}$ and with $\delta_0 = (t_0 -t)~(Dt_0)^{-{1\over 3}}$. The
difference of these exponents is
\beq
{\delta_0\eta_0-\delta\eta\over 2} - {1\over 4}~\bigg({\delta^2\over\eta} -
{\delta^2_0\over\eta_0}\bigg) + {\eta^3-\eta_0^3\over 12} = \Delta~.
\label{fiftytwo}
\eeq
When ${\Delta t\over t} \ll 1$,
\beq
\Delta \simeq \delta(Dt)^{{1\over 3}}~{d\over dt}~\bigg(
{\delta\eta\over 2} - {\delta^2\over 4\eta} + {\eta^3\over 12}\bigg)~.
\label{fiftythree}
\eeq
Using $\delta\propto\eta^2$ as the important values of $\delta$ in
(\ref{fifty}), one easily finds
\beq
\Delta \propto{\eta^5\over t^{{2\over 3}}} \sim \omega_sY ({Y^4\over
t^8})
\label{fiftyfour}
\eeq
which is exactly the size of the error estimated in (\ref{thirtytwo}). This
conclusion is consistent with the estimate ${\Delta\over \omega_sY} =
0({Y^4\over t^8})$ obtained from the heuristic argument based on Eq.
(\ref{twentytwo}).

\section{The very large $Y$ regime}

In the previous section we have exploited the decomposition
(\ref{twentyeight}) of the Green's function in two terms, one (Eq.
(\ref{twentynine})) with product of Airy functions out of phase $(\sim
K(\xi)
K(e^{\mp i\pi}\xi_0))$, and the other in phase $(\sim K(\xi)K(\xi_0)$, Eq.
(\ref{thirty})), corresponding to the exponents
\beq
{\cal E} (\omega, Y) = \omega Y - {3\over 2} \xi^{{3\over 2}} \pm {3\over 2}
\xi^{{3\over 2}}_0~.
\label{fiftyfive}
\eeq
We have shown that, for $Y \ll t^2$, the integral is dominated by Re$
\omega
\simeq \omega_s$, the contribution (\ref{thirty}) being negligible on a
contour with a slight deformation in Re$\omega > \omega_s$. On the other
hand, for $Y \gg t^2$, it becomes profitable to distort the contour in
Re$\omega < \omega_s$, where the contribution (\ref{thirty}) becomes
actually
{\it dominant}, because Re$ \xi^{{3\over 2}}$ may take negative
values.

In fact, for small enough $\omega$ in Im$\omega {>\atop<} 0$, the exponent
with ($-$) sign in Eq. (\ref{fiftyfive}) takes the form
\beq
{\cal E}_\pm \simeq \omega Y \pm {4i\over 3}~{t\omega_s(t)\over D\omega}
\label{fiftysix}
\eeq
with derivatives
\beq
{d{\cal E}_\pm\over d\omega} =  Y \mp {4i\over 3}~{\chi_m\over
b{\sqrt{D}}\omega^2}~,~~~
{d^2{\cal E}_\pm\over d\omega^2} =   \pm {8i\over 3}~{\chi_m\over
b\sqrt{D}\omega^2}~,
\label{fiftyseven}
\eeq
Therefore, there are saddle points of the exponent (\ref{fiftysix}) at
\beq
\omega_\pm = \sqrt{{2\over 3}} (1 \pm i)~\bigg({\chi_m\over b
Y\sqrt{D}}\bigg)^{{1\over 2}}
\label{fiftyeight}
\eeq
with
\beq
{d^2{\cal E}_\pm\over d\omega_\pm^2} = (1 \mp i)~\bigg({3b\sqrt{D}\over 2
\chi_m}\bigg)^{{1\over 2}} Y^{{3\over 2}}~.
\label{fiftynine}
\eeq
Since $(d^2{\cal E}_\pm / d\omega^2_\pm)^{-{1\over 2}} \ll \omega_\pm$ the
gaussian integration, in (\ref{twentyfour}), about the saddle points should
be
an accurate evaluation of the integral. One finds
\beq
g_P(Y;t,t_0) = {t_0\over 3\sqrt{\pi}}~\bigg({3b\over 4\chi_mD^{5/6}
Y}\bigg)^{{3\over 4}}~\bigg[ \cos (\bar\omega \sqrt{Y} + {\pi\over
8})\bigg]
e^{\bar\omega\sqrt{Y}}~,
\label{sixty}
\eeq
with $\bar\omega^2 = {8 \chi_m\over 3b\sqrt{D}}$.

It then appears that the perturbative solution becomes smaller than predicted
by the exponent $\omega_s(k k_0)Y$ (because $\sqrt{Y}\ll \omega_s Y$ for
$Y\gg t^2$) and furthermore it starts oscillating, thus loosing positivity. 
It is no surprise that, at such large $Y$ values, our
perturbative calculation no longer has physical sense. 
 After all, when $Y$ approaches $t^2$ from below it is clear from
(\ref{twentyone}), or (\ref{fifty}), that $\Delta t$ approaches $t$ at which
point one expects the singular potential evident in (\ref{five}) to become
troublesome. That is, when $Y$ approaches $t^2$ the cutoff at $\bar t$ is
clearly necessary.
%Although we have control of the perturbative solution for
%$g_P$ when $Y \ll t^2$ and when $Y \gg t^2$, we have been unable to
%determine
%(perturbatively) $g_P$ when $Y$ is on the order of $t^2$.
Two questions then arise: firstly, whether we can still describe the
behaviour of $g_P$ in the intermediate region $Y~\propto~t^2$; and, secondly,
at which $Y$ values does the non-perturbative Pomeron part really take over.

We are unable to answer either question in detail. However, for the pure Airy
model described by Eq. (\ref{twentytwo}), we can provide a partial resummation
of the corrections to the exponent of relative order $(Y/t^2)^{2n}$ as
follows. Referring to Eq. (\ref{twentyfour}) we first neglect, in the large
$Y$ regime, the term ${\partial ^2 E \over \partial t^2}$ compared to 
$({\partial E \over \partial t})^2$;
then we consider the exponent $E(Y;t,\Delta t)$ around its maximum in $\Delta
t$ so as to neglect its $\Delta t$ derivative, and we take the ansatz
\beq
E(Y;t,\Delta t_{\rm max})=\omega_s(t)Y~f(z),~~z=D\chi_m{Y^2 \over
  t^4}={{\eta}^3 \over {\omega_s Y}}~.
\label{sixty1}
\eeq 
which is supposed to describe the $Y^2/t^4$ dependence. Finally, by replacing
Eq. (\ref{sixty1}) into Eq. (\ref{twentytwo}), we get the non-linear
differential equation
\beq
f(z)+2z f'(z)~=~1+z[f(z)+4z f'(z)]^2~,
\label{sixty2}
\eeq
which is a sort of Hamilton-Jacobi limit of the diffusion equation
(\ref{twentytwo}).

 For $z \ll 1$ ($Y \ll t^2$), the iterative solution to Eq. (\ref{sixty2}) is
 $f(z)=1 + z/3 + 2z^2/3 +...$, which yields the ${\eta^3/3}$ term of
 Eq. (\ref{fifty}) and the first non-trivial correction to it. On the other
 hand, for $z\gg 1$ ($Y \gg t^2$), Eq. (\ref{sixty2}) still makes sense, with
 a solution $f(z) \sim z^{-1/4} + O(z^{-1/2})$, in order to compensate the
 term $1$ in the r.h.s. Therefore, since $z \propto Y^2/t^4$, the exponent in
 Eq (\ref{sixty1}) turns out to be in agreement with the saddle point estimate
 in Eq. (\ref{sixty}). Note, however that while the behaviour (\ref{fifty}) is
 supposed to be universal for $Y \ll t^2$ (Cf. Sec. 6), the resummation in
 Eq. (\ref{sixty2}) and the large $Y$ regime in Eq. (\ref{sixty}) are typical
 of the Airy model, because they involve large $\xi$ properties of
 Eq. (\ref{twentytwo}).

We wish now to compare the large $Y$ behaviour of the perturbative term in
Eqs. (\ref{twentyone}) and (\ref{fifty}) - with its intricate transition to
the behaviour (\ref{sixty}) just discussed - to the non-perturbative Pomeron
term. The latter can only be discussed in a model dependent way. In the Airy
model defined in Eqs (\ref{eight}) and (\ref{nine}) it is straightforward to
show that 
\beq
g_R(Y;t,t_0)~\simeq~ e^{\omega_PY}{\pi t_0 \over \omega_P}\bigg({2b\omega_P
 \over \chi_m''}\bigg)^{2/3}~Ai^2(\xi_P) 
\simeq  \exp[{\omega_PY-{4\over3}
t^{3/2}\bigg({b\omega_P\over {\chi_mD}}\bigg)^{1/2}}],
\label{sixty3}
\eeq
where $\xi_P\sim t(b\omega_P/ \chi_m D)^{1/3}$ and we have
assumed $\omega_P \gg \omega_s(t)$, for definiteness.
It is clear from Eq. (\ref{sixty3}) that the Pomeron term is exponentially
suppressed with respect to the perturbative one for large $t$. However, it
takes over at very large energies, such that
\beq
Y~>~Y_c \simeq {4t^{3/2}\over 3\omega_P}\bigg({b\omega_P\over{\chi_mD}}
\bigg)^{1/2}~,
\label{sixty4}
\eeq
that is, even before the region $Y\simeq t^2$ is reached, as already pointed
out for collinear models in \cite{cc3}.

The above estimate of $Y_c$ is model dependent in several ways. In fact the
weight of the Pomeron term depends on the small-$x$ equation being used
(cf. Sec. (6)) and, within the given equation, on the way the strong coupling
region is smoothed out or cutoff \cite{ccs}. Furthermore, unitarity
corrections may affect it, and there is no consensus on how to 
incorporate
them.  However, the basic qualitative feature underlying all models is that
``tunneling'' to the Pomeron behaviour occurs at some $Y_c < t^2$,
i. e. before the perturbative calculation looses sense, thus insuring
cross-section positivity.

\section{Extension to collinear small-$x$ models}

The above evaluation of perturbative diffusion properties of the Airy model
(Eqs. (\ref{twentyone}) and (\ref{fifty})) can be generalized to other 
small-$x$
models, for which the expression (\ref{twelve}) of the perturbative Green's
function is sufficiently explicit.

For instance, a simple two-pole collinear model \cite{cc3} is provided by the
effective eigenvalue function
\beq
\chi (\gamma,\omega) = {1\over\gamma} + {1\over 1 + \omega-\gamma}~,
\label{sixtyone}
\eeq
where we have introduced the $\omega$-shift \cite{ccs2}, but no further
subleading terms. The corresponding Green's function is defined by \cite{mt}
\beq
[\omega - \chi ({1\over 2} + \partial_t,\omega){1\over bt}]~ g_\omega(t,t_0)
=
\delta (t-t_0)~,
\label{sixtytwo}
\eeq
which basically provides a second order differential equation for
$g_\omega$.
Therefore, the solution is of type in Eq. (\ref{two}), which, for $t > t_0 >
\bar t$, reads
\bea
g_\omega (t,t_0) &=& {1\over\omega}~\delta (t-t_0) + {1\over \omega^2 bt_0}~
W_{{1\over b\omega},{1\over 2}} \bigg[(1+\omega)t\bigg] \times \nonumber \\
&& \times \bigg[ W_{-{1\over b\omega}, {1\over 2}}(-(1+\omega)t_0) +
S(\omega)
W_{{1\over b\omega},{1\over 2}}((1+\omega)t_0)\bigg]~.
\label{sixtythree}
\eea
Here  $W_{k,m}(z)$ is the Whittaker function, satisfying the Coulomb-like
differential equation
\beq
\bigg[ -{d^2\over dz^2} - {k\over z} + {m^2-{1\over 4}\over z^2} + {1\over
4}\bigg]~W_{l,m}(z) = 0
\label{sixtyfour}
\eeq
and its ``irregular" counterpart ``$W_{-k,m}(z)$" is more precisely defined
by
the real analytic continuation
\beq
W_{-k}(-z) = {1\over 2}~\bigg(e^{i\pi k} W_{-k} (e^{i\pi}z) + e^{-i\pi k}
W_{-k}(e^{-i\pi}z)\bigg)~.
\label{sixtyfive}
\eeq
Such solutions admit also the $\gamma$-reresentation \cite{ccs}
\beq
W_{\pm{1\over b\omega},1/2}\bigg(\pm (1+\omega)t\bigg) = \int_{C_R,C_I}
{d\gamma\over 2\pi i}~~e^{(\gamma - {1\over 2}t - {1\over
b\omega}~X(\gamma,\omega)}
\label{sixtysix}
\eeq
where $C_R(C_I)$ is the contour for the regular (irregular) solution shown
in
Fig. 2b, and $\chi (\gamma,\omega ) = {\partial\over\partial\gamma} X(\gamma ,
\omega )$.

The energy dependence of the perturbative function $G_P(Y;t,t_o)$ in Eq.
(\ref{twelve}) was already investigated in Refs. \cite{ccs},\cite{mt} and is
characterized by various regimes.
\begin{enumerate}
\item In the collinear limit $t \gg t_0 > 1$, and $Y \gg 1$, $G_P(Y;t,t_0)$
is
dominated by the customary anomalous dimension saddle point, which exists in
the energy region 
\beq
\sqrt{{\log t/t_0\over bY}} \simeq \omega > \omega_s(t_0) \gg
\omega_s(t)~,~~~(Y\equiv \log {s\over k^2})~,
\label{sixtyseven}
\eeq
where $\omega_s(t) = {4\over (1+\omega_s)bt} \simeq {4\over bt}$ is the
saddle
point exponent.
\item In the diffusion region $\Delta t = t - t_0 \ll t$, and $Y \gg t \gg
1$,
the important $\omega$ values drift towards $\omega_s(t)\simeq
\omega_s(t_0)$,
and the asymptotic behaviour of the Green's function matches a properly
chosen
Airy model.
\end{enumerate}

In fact, if $t$ and ${1\over\omega}$ are both large, but $\omega -
\omega_s(t)\simeq\omega - \omega_s(t_0)$ is small, the phase in Eq.
(\ref{sixtysix}) is finite only if $(\gamma -{1\over2})\sim
(b\omega)^{{1\over
3}}$ is small also. This means that we are probing the region close to the
minimum of $\chi(\gamma)$, in such a way that
\beq
\bigg({2b\omega\over \chi_m''}\bigg)^{{1\over 3}}~~\bigg(t - {\chi_m\over
b\omega}\bigg) = \xi = \bigg({t
\omega_s(t)\over\sqrt{D}\omega}\bigg)^{{2\over
3}}~~ \bigg({\omega\over\omega_s(t)} -1\bigg)
\label{sixtyeight}
\eeq
is finite. This is precisely the region where the $W$'s  become
asymptotically Airy functions, with parameters
\beq
\chi_m = {4\over 1+\omega_s}~,~~~ D = {\chi_m''\over 2\chi_m} = {4\over
(1+\omega_s)^2}~,~~~ \omega_s(1+\omega_s) = {4\over bt}~.
\label{sixtynine}
\eeq
For this reason, the linearized expression (\ref{fifteen}) holds in the
present case also, under the same conditions.

The above argument shows that, in the region $t\ll Y \ll t^2$, and $(t-t_0)
\ll t$, the diffusion corrections of Eq. (\ref{nineteen}) are valid for the
two-pole model also, and for any truncated model where the
$\gamma$-representation (\ref{sixtysix}) holds for both the regular and the
irregular solution.

Of course, for $Y \gg t^2$, the important $\omega$ values become much
smaller
than $\omega_s(t)$, and the phases in Eq. (\ref{sixtysix}) became stationary
at $\chi(\bar\gamma,\omega) = 0$. For the Airy model this occurs at $\bar\nu
= \bar\gamma - {1\over 2} \sim \pm{i\over\sqrt{D}}$, and provides the
asymptotic behaviour used in Eq. (\ref{fiftysix}). For the two-pole model,
instead, $\bar\gamma$ drifts to $\pm i\infty$, and this provides the
asymptotic formulas
\bea
g_\omega &\sim& \exp \bigg(\pm {2\pi i\over b\omega}\bigg)~, \nonumber \\
g_P(Y;t,t_0) &\sim& \exp \bar\omega\sqrt{Y}~\cos(\bar\omega \sqrt{Y} +
{\pi\over 8})~,
\label{seventy}
\eea
with $\bar\omega = {4\pi\over b}$. Although the precise exponent is
different,
the qualitative behaviour (\ref{seventy}) is the same as for the Airy model.
This follows from the analogous structure of the $\gamma$-representation
(\ref{sixtyfive}), with its $\exp({{\rm const}\over\omega})$ behaviour.

In conclusion, in the range $t\lappeq Y\ll t^2$, the perturbative calculation
shows a universal behaviour, described by Eq. (\ref{twentyone}) or
Eq. (\ref{fifty}) , where the only model dependence  lies in the parameters
$\chi_m$ and $D$, describing the hard Pomeron and the diffusion
coefficient. For $Y\gappeq t^2$ on the other hand, the perturbative behaviour
is more model dependent and finally starts oscillating, thus loosing physical
sense. 

At some intermediate value $Y=Y_c$ ($t\ll Y_c\ll t^2$), the non-perturbative
Pomeron takes over. We do not have a reliable model for such transition. Is it
an abrupt ``tunneling'' effect \cite{cc3}, as in the two-pole model (with
$\omega_P\gg \omega_s(t)$), or is it instead mediated by a long diffusion
regime, as perhaps expected in unitarized models ($\omega_P \ll 1$)? This is
an open question, which deserves further investigation.


\begin{thebibliography}{99}
\bibitem{bfkl}
 L.N. Lipatov,{\it Sov. J. Nucl. Phys.} {\bf 23} (1976) 338;
E.A. Kuraev, L. N. Lipatov and V. S. Fadin {\it Sov. Phys. JETP} {\bf 45} 
(1977) 199; 
Ya. Balitskii and L. N. Lipatov, {\it Sov. J. Nucl. Phys.}{\bf 28} (1978) 822.
\bibitem{fl}
V. S. Fadin and L. N. Lipatov, \PL {\bf 429B} (1998) 127 and references
 therein. 
\bibitem{cc1}
G. Camici and M. Ciafaloni, \PL {\bf 412B} (1997) 396 and {\bf 430B} (1998)
 349. 
\bibitem{ck}
J. C. Collins and J. Kwiecinski, \NP {\bf B316} (1989) 307.
\bibitem{cc2}
G. Camici and M. Ciafaloni, \PL {\bf B395} (1997) 318.
\bibitem{km}
Yu. V. Kovchegov and A. H. Mueller, \PL {\bf B439} (1998) 428.
\bibitem{abb}
N. Armesto, J. Bartels and M. A. Braun, \PL {\bf B442} (1998) 459.
\bibitem{ll}
E. M. Levin, \NP {\bf 445B} (1999) 481.
\bibitem{cc3}
M. Ciafaloni, D. Colferai and G. P. Salam, {\it JHEP} {\bf 9910} (1999) 017 and
 {\bf 0007} (2000) 054.
\bibitem{ccs}
M. Ciafaloni, D. Colferai and G. P. Salam, \PR {\bf D60} (1999) 114036.  
%\bibitem{uu}
\bibitem{mt}
M. Taiuti, {\it University of Firenze Thesis} (2000), unpublished.
\bibitem{ccs2}
G. P. Salam,{\it JHEP} {\bf 9807} (1998) 19;\\ 
M. Ciafaloni, D. Colferai, \PL {\bf 452B} (1999) 372.
%\bibitem{nn}

\end{thebibliography}
\end{document}